# An evaluation strategy of the thermal conductance of semiconductor interfaces using ultraviolet light.

Dian Li[1], Joseph Feser[1]

[1] *Department of Mechanical Engineering, University of Delaware, 19716, DE, USA*

# ABSTRACT

In the previous studies, the ultraviolet light thermoreflectance (UV-TDTR) signal from bulk semiconducting samples cannot be well explained by a thermal models based on the assumption that the heat is both absorbed and probed near the surface. A thermoreflectance (TDTR) technique was developed to directly excite semiconductors using UV-TDTR. At $\lambda = 400nm$, the photon energy is much greater than most semiconducting bandgaps, potentially allowing semiconducting transducers to absorb light within about 10 nm of the surface, potentially enabling direct measurements of semiconductor-semiconductor interfaces. The thermoreflectance coefficient for some direct-bandgap semiconducting materials was measured. Thermal transport models were fitted to thermorefectance data collected from the system. The differences in the signal for semiconductors with different doping and nanostructure features were also studied, which should change the recombination rate. No significant changes to the signal were observed.

# 1. Introduction

The physical propertied of metal materials receive a lot of research interests [2, 3, 4, 5, 6, 7, 8, 9, 10, 11]. In metals, as an essential source of heat conduction, free electrons carrying heat vibrate against the neighboring electrons, thus exchanging their energy [12, 13, 14, 15, 16, 17]. The hotter free electrons become, the faster they propagate and vibrate, which increases the chance for hot electrons scatter to all directions and exchange energy with other electrons with lower energy [1, 18, 20, 21, 22]. In addition, free electrons are not the only factor contributing to heat conduction [19, 24]. The vibration of lattice also helps carry energy among temperature gradient in solids [15, 25, 26, 27]. In all solids, as long as there are atoms, the vibration of phonons plays an indispensable and important role in thermal transport [27, 29, 30, 31, 32]. Therefore, it is simple to say that for solids such as metals or semiconductors, thermal transport may be separated into two independent parts, electron thermal transport and phonon thermal transport.

In the previous studies, investigation on the thermal boundary conductance (thermal interface conductance) of the interface between epitaxial semiconductors is still lacking [33, 34, 35, 36]. Normally, metallic films are used as the top layer for TDTR measurements (aka the 'transducer') because of their ability to absorb laser heat in a thin layer near the surface (typically tens of nm), provide a reasonably large thermoreflectace coefficient in many cases (typically $10^{-5} - 10^{-4}$ K$^{-1}$, and high thermal conductivity. As a result, most studies of interface conductance have been for metal-semiconductor interfaces where the metal transducer is the top layer; there are few experimental works probing the heat transfer happening at the interface of semiconductor/semiconductor. It may be assumed that the interface heat conduction happening at such interface is similar to that in normal metallic interfaces, such as superlattice interfaces. However, this has been proved not correct. Several works reported that the interaction dueto the interfacial layers does exist and has an influence on thermal transport across them. Garg and Chen [37] firstly contributed to the work of finding the real relationship between the boundary interface thermal conductance and the heat deposition. Few experimental studies of thermal interface conductance exist for individual epitaxial semiconductor/semiconductor interfaces at room temperature and above. In our view, this is because TDTR is optimized for metallic light absorption. Thus, most thermal interface conductance "measurements" have not been done for individual interfaces, but rather have been extrapolated from superlattice measurements assuming that there is no interaction between adjacent interfaces. The last assumption is incorrect in many cases. In the work of Luckyanova, it was found that the thermal conductivity of the AlAs/GaAs superlattice configuration multilayers increase linearly with the total thickness, which may be explained by the coherent phonon heat conduction process. Other studies have attempted to circumvent this hole in the TDTR capabilities by using a tri-layer stack (thin metal-semiconductor-semiconductor) but since the primary sensitivity then becomes to the metal-semiconductor interface, this leads to high error bars for the semiconductor-semiconductor thermal interface conductance.

In this study, a new experimental variant of TDTR was proposed. It allows direct and single-interface measurements of semiconductor/semiconductor thermal interface conductance. This method is based on creating an all-ultraviolet (UV) TDTR variant designed to use semiconducting layers as both the optical absorber and temperature probe, allowing sensitivity to interface conductance at non-metal interfaces. It has the advantage of enhancing the absorption coefficient, such that optical absorption would occur in most cases in tens of nanometers.

## 2. Methods

2.1. Newly developed UV-TDTR method

An all-ultraviolet (UV) TDTR variant was created, which uses semiconducting layers as both the optical absorber and temperature probe, allowing to measure interface conductance at non-metal interfaces. Using UV-TDTR has the advantage of enhancing the absorption coefficient, such that optical absorption would occur in most cases in tens of nanometers. TDTR sensitivity to interface conductance requires absorption distances smaller than $\delta \ll \kappa/G_{int}$ for good sensitivity, where $\kappa$ is the thermal conductivity of the transducer and $G_{int}$ is the thermal interface conductance to be measured. For a high thermal conductivity semiconductor with high thermal interface conductance (100 W/m-K and 1000 W/m$^2$-K respectively, say) this gives 100 nm as the upper limit of allowable light absorption depth. In most semi- conductors at conventional TDTR wavelengths (800nm), this condition cannot be met (see Table 1). On the other hand, in many semiconductors with bandgaps from 1-3eV, using second harmonic generation (SHG) from a standard femtosecond laser (800nm→400nm wavelength) would achieve sufficient absorption depth; using a typical sum-frequency generation (800nm→267nm wavelength) would further enable measurements on wide bandgap semiconductors with gaps up to 4.5eV, including some technologically important ones such as SiC (3.0eV) and GaN (3.4eV).

At SHG wavelengths, most of the optics used for the UV-TDTR could be analogous to conventional TDTR: (1) a single UV-laser created using SHG from a $\lambda$=800nm source could be split according to polarization into a pump and probe beam, (2) with the pump beam traveling through a conventional electro-optic modulation, and (3) beam combination and pump rejection after the sample reflection handled by a second polarizing beam splitter (PBS). A combination of geometric blocking and audio frequency chopping of the probe laser could be used to reject the 1:1000 leaked pump light that makes it past the second PBS upon return from the sample. Sun et al. [14] has recently shown that, using audio-frequency chopping rather than traditional two-tint pump beam reject can even enhance signal-to-noise ratio by up to 4-fold by reducing wavelength-jitter induced noise and increasing the utilization of laser-light. By using harmonic ultrafast mirrors, we believe it will even be possible to switch quickly betweenconventional and UV TDTR to maintain concurrently capabilities.

2.2. 400 nm Ultraviolet-TDTR system

To meet the experimental requirements of the thermal transducer selection un- der ultraviolet wavelength, a revised system of the current 785 nm TDTR system in thelab is developed dedicated for ultraviolet time domain thermoreflectance (UV-TDTR) Figure 1. In this system, a few components are added to the existing TDTR system to produce laser at ultraviolet wavelength while several other components are replaced or removed to ensure the best measurement quality of UV-TDTR. In the UV-TDTR system, the initiated laser from Ti:Sapphire has a center wavelength of 800 nm. This is controlled by adjusting the BRF micrometer of the Ti:Sapphire. This wavelength is needed to produce the 400 nm laser the UV experiment. Different from infrared TDTR, the 800 nm laser beam goes through second harmonic generation before being split into the pump and the probe beams. There isa set of two lenses which have the same focal length, $f$. They are placed $2f$ apart with a nonlinear optical crystal in between. This combination lets a light coming through them be focused on the center between them and expanded back to its normal shape an recollimated. At the center point, a nonlinear Barium Borate (BBO) crystal is placed to initiate second harmonic generation. The reason that this crystal is physically required at the focal point is because that the transformation efficiency of frequency doubling in the pump laser beam is dependent on the intensity of the pump laser. The relationship between the intensity of the second harmonic wave and the pump beam intensity is

$$I_s = \gamma I_p^2 \tag{1}$$

where $I_s$ refers to the second harmonic wave intensity and $I_p$ stands for the pump beam intensity. $\gamma$ is a factor that is related to the shape, non-linearity and material type of the crystal. Thus, we need to place the BBO crystal at the focal point to gain highest efficiency of second harmonic generation. The generated second harmonic pump beam power can be as high as 120 mW when the 800 nm laser power output is 1.4W. A UV filter which filters the infrared 800 nm laser further provides the laser needed for the UV-TDTR experiment. Notably, during the process of second harmonic generation, sum frequency generation also happens which produce the invisible middle ultraviolet laser at 267 nm. In order to find its influence on the sample's reflectance variation, there is a bandpass filter (Thorslabs FGB25M 315 - 445 nm & 715 - 1095 nm) added to further filter the third harmonic laser. As we want to create a spot size as small as possible to the sample, another beam expander (BE02-05A) is placed at the path to increase the beam size to as large as $D= 2.7$mm, which is the diameter of the inlet path of the EOM. Also, it can change the laser Gaussian profile so the beam size of the pump laser at magnification lens while the delay stage is moving will not change by a large amount.

As for the measuring path for UV-TDTR, there are some minor modifications to the normal TDTR system. First, an aperture is placed between the photo diode lens and UV nonlinear beam splitter to block the leaked pump beam. We set the heightof the probe beam at 12.7 mm (5 inches) while the pump beam 3 mm lower thanthe probe beam height. The magnification lens is also centered at 12.7 mm height. After the reflection, the reflected probe beam remains at the same height while the reflected pump beam is at 13.0 mm from the stand. The aperture is centered at 12.7 mm with a 4 mm diameter opening. This placement has blocked the leaked pump beam from passing through the aperture and thus the effect of the pump beam leakage is removed. For the photo diode, we use a low wavelength one (Thorslabs DET25K) for better absorption of 400 nm lights. This GaP detector has an operation wavelength range of 150 - 500 nm and has a responsivity of 0.11 at 400 nm, larger than 0.08 of the Si detectors for Infrared TDTR. The RLC circuit is not installed into the UV-TDTR system as the bandwidth of the photodiode (6.4 MHz) is less than 12.6 MHz, and the corresponding capacitance is higher and not suitable for a high Q-factor resonant filter.To compensate for the lack of RLC amplification, we use an additional stage of preamplification when the second channel of the preamplifier is connected to output a factor of $5 \times 5 = 25$ times of amplified signal (this also amplifies the first stage of amplifier noise). We have applied a sinusoidal wave with amplitude of 0.6 V and 3 V offset from the function generator at 6 MHz modulation frequency. This must meet the requirement of the photo detector. In addition to the changes to the system, we have replaced retroreflector and beam splitters with the same components dedicated for ultraviolet wavelengths so the loss of reflection along the path is reduced. It is crucial to choose a good retroreflector since there are three reflections on the delay stage. Using aluminum in- stead of our usual protected gold increased the overall reflectance to $R^3_{Al}(400nm)=0.79$ from $R^3_{Au}(400nm)=0.06$.

3.3. UV-TDTR thermal transducer choice

Ideally, for a single target film, when shot with a UV-laser, the temperatureinduced change in reflectance is expected to give a signal

$$V_{in} \propto -\frac{V_{in}}{V_{out}} \propto t^{-\frac{1}{2}} \quad 1 \tag{2}$$

Eq. (2) stems from the pulse surface-heating temperature profile on a semi- infinite plate surface where T is determined by

$$T(x,t) = \frac{Q_0}{\sqrt{\pi k C t}} \exp\left(-\frac{x^2}{\sqrt{4\alpha t}}\right) \tag{3}$$

where $Q_0$ is the energy deposited by single laser hating pulse and α is the thermal diffusivity. For a good thermal transducer at UV wavelength, when it absorbs the pulsed heating from the laser, the thermoreflectance in response to its surface temperature has its TDTR thermal profile curve over time delay should be parallel to $t^{-1/2}$ ('ideal' curve). Thus, as a control measurement, we have attempted to show that for a single layer material (i.e. a substrate only) we obtain this profile, and thus a thermal signal.

3. Results and Discussion

We have applied UV-TDTR scan on various semiconductor substrate samples to characterize the thermoreflectance coefficient and to check that the signals are indeed thermal in nature (i.e. $V_{in} \propto t^{-1/2}$). The laser output powers we use are 11.58 mW for the pump beam and 18.39 mW for the probe beam. Because of the relatively low efficiency of the second harmonic generating 400 nm from the 800 nm laser, we use a 10X lens to get higher signal on the measurements. Both Figure 3 and Figure 4 illustrate the thermal response of pulsed laser at 400 nm wavelength. From Figure 2, it shows that the thermal profiles of the listed semiconductors tend to have the t−1/2 profile, which is expected. Based on the time delay of the absorption processes at each sample and the acoustic speed of these semi- conductors, we can calculate the actual thermal absorption depth of the listed samples.

The absorption processes of all the listed samples roughly happen at t= 25 ps. Referring to the acoustic speed of GaAs and GaSb, we conclude that the actual UV- pulse thermal absorption depth of GaAs and GaSb are 59 nm and 49.6 nm. For a lack of sources of the acoustic speed, we could not determine the actual thermal absorption depth of ErSb:GaSb and ErSb:InGaSb. However, if we assume the acoustic speed of both ErSb:GaSb and ErSb:InGaSb is 4 nm/ps, then the assumed actual thermal absorption depth at 400 nm UV laser is 50 nm.

We see some evidence for various semiconductors that meets the relationship in Eq.(2) that we expected for UV-TDTR, but only in the range from 100-1000ps. However, the fact that many of the signals cross through zero at long time delay indicates that these signals are not entirely thermal in nature, meaning that the listed semiconductors might not in fact be appropriate for UV-TDTR thermal transducers. The actual absorption depth of the listed samples is much greater than the values in Table 1. Based on Table 1, the time delay point corresponding to the thermal absorption depth for GaAs, GaSb, and ErSb:GaSb & ErSb:InGaSb are 6.34 ps, 8.06 ps and 10 ps(assumed). In Figure 4, the difference between the actual experimental absorption points and those from Table 1 on UV-TDTR scan have great difference. Therefore, it is concluded that for GaAs and GaSb, the heat absorption process caused by laser pulse heating does not happen in the depth where the heat is deposited. Additionally, it was observed that the UV laser brings about some irreversible effect on the semiconductors. When shooting the laser with stable power to the semiconductors, the detect voltage of the TDTR system steadily drops from time to time.

In Figure 5, the first 600s of the detector voltage of GaAs change is recorded. It is obvious that the detector voltage drops as time passes exponentially. More surprisingly, this effect is irreversible. We turned off the laser after it caused such effect and waited for 20 minutes. Then the laser is turned on again to irradiate the same region of the sample. The detector voltage was still the value as it ended up with for the first irradiation. We make sure that the detector voltage drop in this case is notdue to the fluctuation of the probe beam by measuring the probe beam power for the same amount of time. It turned out that the probe beam power output kept stable within a range of minimal changes. Therefore, the probe beam power itself has been eliminated from the cause of this phenomenon.

To investigate this problem, we have made some hypotheses. First, we assume this detector voltage drop may stem from the effect brought about by impurity laser (laser in other wavelengths below 400 nm) from the UV beam. The reason we propose this case is that during SHG process of 800 nm laser, there may be 267 nm UV light produced because of sum frequency generation (SFG) which could simultaneously while SHG happens. As a result, there may be 267 nm UV light emitted along with the 400 nm UV laser from the nonlinear crystal. To confirm whether the SFG caused the detector voltage drop, we applied a colored bandpass wavelength filter (Thorslab FGB25M) right following the nonlinear crystal, which only passes UV laser with the wavelength between 315 nm and 445 nm. Once this filter is applied, wesaw an approximately 30 % power drop compared to the previous unfiltered UV light. However, after the filter is applied, the voltage drop persisted. Then the cause because of the laser in other wavelengths is excluded.

It is worth mentioning that this voltage drop should not be due to the free electron movement, too. Because if the free electron distribution owing to the irradiationof the UV laser leads to the voltage drop of the photo detector, then the voltage willresume to the original value once the laser is turned off for some time and turned onagain. This means the movement of free electrons will not result in permanent damage to the sample, which was observed.

4. Conclusion

During the UV-TDTR scan of semiconductor samples, the UV laser is not depositing the pulse heating at the surface or near surface region of the semiconductor samples. Rather, from TDTR curves, it is concluded that, in this case, the heat is concentrated in the region much deeper than the absorption layer. This brings a challenge to the candidate semiconductors as UV thermal transducers as the thickness minimum must be large. This means depositing heat in a thin region nearthe surface (usually less than 20 nm) in these semiconductors is no longer feasible, making these non-ideal as thermal transducers.

**Figure Captions**

Table 1. Optical Absorption Depth for Some Common Materials for wavelengths corresponding to the proposed ultraviolet TDTR: SHG=400nm, SFG=267nm. Conventional TDTR uses $\lambda$=800nm for at least one laser beam.

| Material | $\alpha^{-1}$ ($\lambda$=800nm) | $\alpha^{-1}$ ($\lambda$=400nm) | $\alpha^{-1}$ ($\lambda$=267nm) | Ref(s) |
|---|---|---|---|---|
| Al | 7.6nm | 6.6nm | 6.7nm | [1] |
| GaAs | 570nm | 15nm | 6.0nm | [18] |
| AlAs | $\infty$ | 3$\mu$m | 7nm | [14, 16, 18, 19] |
| GaSb | 142nm | 16nm | 5.9nm | [23] |
| InAs | 200nm | 16nm | 10nm | [23] |
| InP | 302nm | 18nm | 7.3nm | |
| GaN | > 50$\mu$m | > 5$\mu$m | 60nm | [13, 23, 24] |

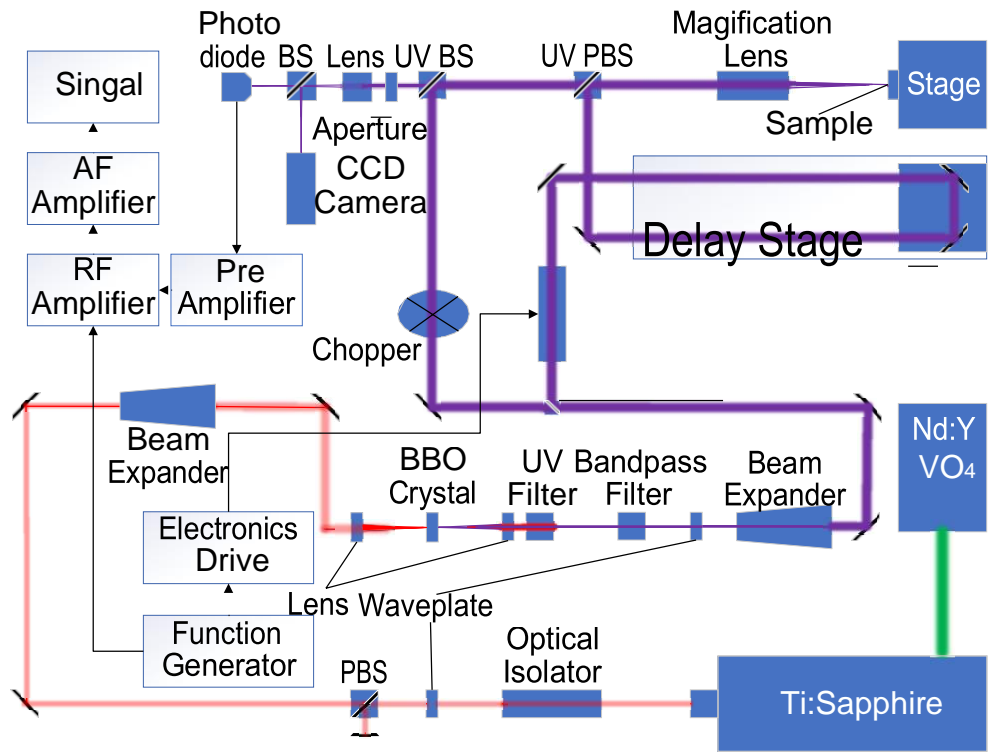

Figure 1. 400 nm Ultraviolet TDTR laser system optical and electric layout

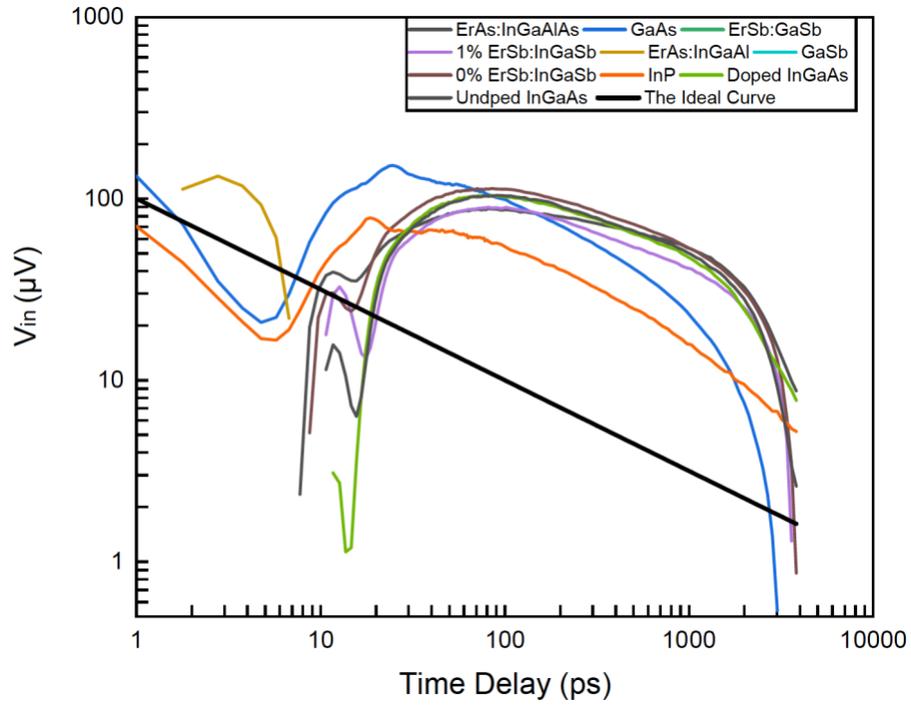

Figure 2. UV-TDTR In-Phase scan of a range of semiconductors from 0.8 ps to 1.6ns and the ideal curve as reference from 1 ps to 3.6 ns time delay in log-log scale.

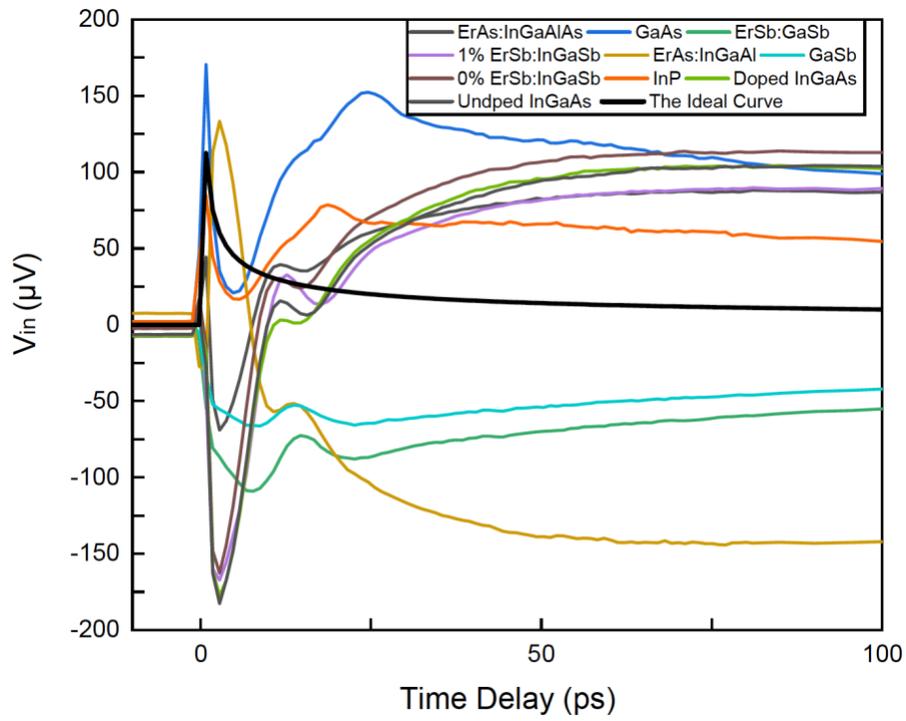

Figure 3. UV-TDTR In-Phase scan of a range of semiconductor at the first 100 ps time delay and the ideal curve as reference in normal linear scale

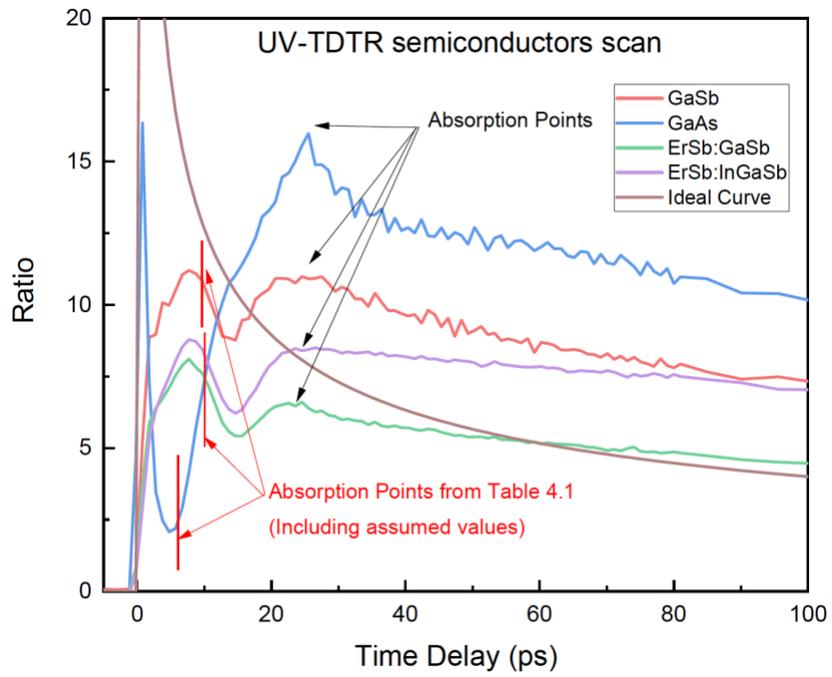

Figure 4. The transit from the unknown UV-TDTR profiles to a normal temperature decay of the tested semiconductor samples.

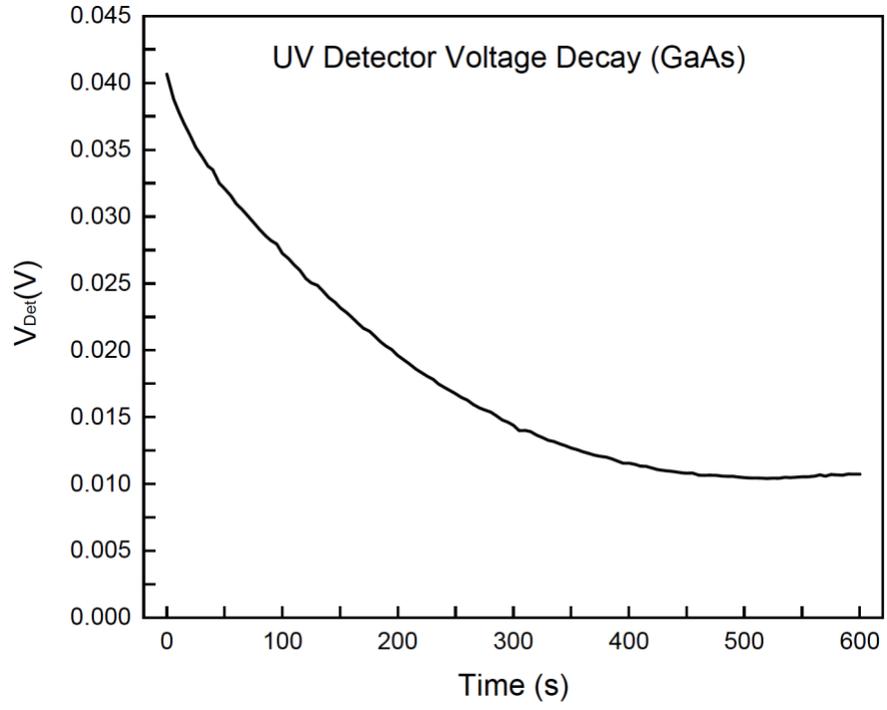

Figure 5. Detector voltage of GaAs at UV-laser for the first 600 s